\documentclass[aps,prl,a4paper,twocolumn,groupedaddress,english,longbibliography]{revtex4-1}

\usepackage{amssymb}
\usepackage{amsmath}
\usepackage{mathtools}
\usepackage{graphicx}
\usepackage{xcolor}

\definecolor{darkblue}{rgb}{0.1,0.2,0.6}
\definecolor{darkred}{rgb}{0.8,0.1,0.2}
\usepackage[colorlinks,
            citecolor=darkblue,
            linkcolor=darkred,
            urlcolor=darkblue]{hyperref}

\usepackage[all]{hypcap} 

\makeatletter
\adddialect\l@English\l@english
\makeatother

\begin{document}

\title{Many-body localization as a large family of localized ground states}

\author{Maxime Dupont}
\affiliation{Laboratoire de Physique Th\'eorique, IRSAMC, Universit\'e de Toulouse, CNRS, UPS, France}
\affiliation{Department of Physics, University of California, Berkeley, California 94720, USA}
\affiliation{Materials Sciences Division, Lawrence Berkeley National Laboratory, Berkeley, California 94720, USA}
\author{Nicolas Laflorencie}
\affiliation{Laboratoire de Physique Th\'eorique, IRSAMC, Universit\'e de Toulouse, CNRS, UPS, France}

\begin{abstract}
Many-body localization (MBL) addresses the absence of thermalization in interacting quantum systems, with non-ergodic high-energy eigenstates behaving as ground states, only area-law entangled. However, computing highly excited many-body eigenstates using exact methods is very challenging. Instead, we show that one can address high-energy MBL physics using ground-state methods, which are much more amenable to many efficient algorithms. We find that a localized many-body ground state of a given interacting disordered Hamiltonian $\mathcal{H}_0$ is a very good approximation for a high-energy eigenstate of a parent Hamiltonian, close to $\mathcal{H}_0$ but more disordered. This construction relies on computing the covariance matrix, easily achieved using density-matrix renormalization group for disordered Heisenberg chains up to $L=256$ sites.
\end{abstract}

\maketitle

\noindent\textit{Introduction.} The mutual effect of disorder and interactions in quantum many-body systems can lead to fascinating phenomena beyond single-particle Anderson localization~\cite{anderson1958,evers2008}. In that respect, many-body localization (MBL) is a key topic which has recently triggered huge activity~\cite{gornyi_interacting_2005,basko_metalinsulator_2006,altman_universal_2015,nandkishore_many-body_2015,abanin_recent_nodate,alet_many-body_2018}. While MBL physics addresses ergodicity and thermalization properties of highly excited states, it is legitimate to ask whether zero-temperature physics may have some connections with MBL.
In this sense, the so-called Bose-glass  phase~\cite{giamarchi1987,giamarchi1988,fisher1989}, which traces back to $^{4}$He in porous media~\cite{reppy1992}, describes an {\it{interacting and localized}} zero-temperature bosonic fluid lacking superfluid coherence  in a  disordered potential. Such an interacting-disordered ground state (GS) has been reported ever since in various contexts~\cite{nohadani2005,fallani_ultracold_2007,hong2010,sacepe2011,yu2012,zheludev2013,kondov_disorder-induced_2015,dupont2017} and theoretically intensively investigated, especially regarding disorder-induced quantum phase transitions~\cite{gurarie2009,altman_superfluid-insulator_2010,alvarez2015,ng2015,ristivojevic2012,hrahsheh_disordered_2012,doggen2017}.

Beyond GS properties, it is now broadly accepted that in one dimension strong enough disorder leads to MBL at {\it{any}} energy, breaking the so-called eigenstate thermalization hypothesis (ETH)~\cite{deutsch1991,srednicki1994,rigol2008}.
Interestingly, MBL is associated with an emergent integrability~\cite{serbyn2013,huse2014,imbrie2016,imbrie2016_,rademaker2017,imbrie2017} and area-law entanglement at any energy density~\cite{bauer2013,kjall2014,luitz2015,lim2016,khemani2017}, while it is the usual hallmark of GS of short-range Hamiltonian~\cite{hastings2007,eisert2010,laflorencie2016}. Overall, this makes MBL states look \textit{very like} GS.

In this Rapid Communication, building on this simple idea, we ask whether an arbitrary MBL state could also be the GS of another Hamiltonian. This question falls in the more general following problem~\cite{qi2017,garrison2018}: Given a single eigenstate, does it encode the underlying Hamiltonian? The answer seems positive for any eigenstate of a generic local Hamiltonian~\cite{qi2017} but also for disordered eigenstates, provided they satisfy ETH~\cite{garrison2018}. However, we argue in the following that this statement \textit{no longer} holds for MBL. Precisely, we find that in the limit of infinitely large systems, a localized Bose-glass GS also corresponds to a MBL excited state of a different Hamiltonian that differs only by its local disorder configuration. We also provide numerical evidence that the distinction between localized GS and MBL excited states cannot be made by any set of local or global measurements. Our results are supported numerically by standard exact diagonalization (ED)  for small system sizes and using the density-matrix renormalization group (DMRG) algorithm~\cite{white1992,white1993} for larger systems, up to $L=256$ lattice sites.

We consider the paradigmatic random-field spin-$1/2$ Heisenberg chain, governed by the Hamiltonian
\begin{equation}
    \label{eq:ham}
    \mathcal{H}_0=J\underbrace{\sum_{j=1}^{L-1} \mathbf{S}_j\cdot\mathbf{S}_{j+1}}_{\coloneqq O_0}+\sum_{j=1}^{L} h_j\underbrace{S^z_j}_{\coloneqq O_j}
\end{equation}
with $L$ lattice sites. Open boundary conditions are used for DMRG efficiency, and the antiferromagnetic  coupling $J$ is set to unity in the following. The total magnetization $S^z_\mathrm{tot}=\sum_j S^z_j$ is a conserved quantity of the Hamiltonian and we work exclusively in the $S^z_\mathrm{tot}=0$ sector. The random variables $h_j$ are drawn from a uniform distribution $[-h, h]$. The GS of this model is known to be of the Bose-glass type for any $h\neq 0$
~\cite{giamarchi1987,giamarchi1988}. At higher energy,  a finite amount of disorder $h_c\gtrsim 3.7$ is necessary to eventually move from an ETH to a fully MBL regime~\cite{luitz2015,villalonga2018}.\\
\\
\textit{Covariance matrix and Hamiltonian reconstruction.} We base our work on the ``eigenstate-to-Hamiltonian construction'' method~\cite{qi2017,chertkov2018}. It takes as an input a wave function $|\Psi_0\rangle$,  eigenstate of the Hamiltonian~\eqref{eq:ham} for a given disorder configuration, and a target space of Hamiltonians. We constrain it to have the same form as the original one, i.e., $\tilde{\mathcal{H}}=\tilde{J}O_0 + \sum_{j=1}^L \tilde{h}_j O_j$. Our goal is to find a set of parameters, represented as a vector $\boldsymbol{p}=[\tilde{J},\tilde{h}_1,\tilde{h}_2,...]^T$, for which the input state $|\Psi_0\rangle$ is an eigenstate, beyond the trivial case $\pm\mathcal{H}_0$. To achieve this, the central object is the covariance matrix $\mathsf{C}_{ij} = \langle O_i O_j\rangle - \langle O_i\rangle\langle O_j\rangle$,
of linear size $L+1$ and with the expectation values measured over $|\Psi_0\rangle$. From this definition, one readily shows that the covariance matrix can be used to compute the energy variance of the input state with respect to a Hamiltonian $\tilde{\mathcal{H}}$ in the target space and whose parameters are encoded in $\boldsymbol{p}$,
\begin{equation}
    \label{eq:cov_matrix}
    \sigma^2\Bigl[\tilde{\mathcal{H}},|\Psi_0\rangle\Bigr] = \langle\tilde{\mathcal{H}}^2\rangle-\langle\tilde{\mathcal{H}}\rangle^2=\boldsymbol{p}^{T}\,{\bf {\mathsf{C}}}\,\boldsymbol{p}^{\vphantom{T}} \geq 0.
\end{equation}
If $\boldsymbol{p}$ is an eigenvector of the covariance matrix $\bf {\mathsf{C}}$ with zero eigenvalue, the set of parameters contained in $\boldsymbol{p}$ defines a parent Hamiltonian $\tilde{\mathcal{H}}$ for which the initial input state $|\Psi_0\rangle$ is precisely an eigenstate. We note and sort in ascending order the eigenvalues of ${\bf {\mathsf{C}}}$, $e_1\leq... e_j...\leq e_{L+1}$, with corresponding eigenvectors $\boldsymbol{p}_j$.

In practice,  we ask whether a localized GS $|\Psi_0\rangle$ can also be an excited state of another Hamiltonian $\tilde{\mathcal{H}}$. There are two reasons for this, and the first one is concerned with the density of states of the Hamiltonian~\eqref{eq:ham}. While its GS is unique, the density of states at high energy is exponentially large, which makes it very unlikely to be able to connect each excited MBL eigenstate to a single GS. Second, it is  numerically much more efficient to work with a GS for the input state $|\Psi_0\rangle$ since its computation is not restricted to ED and hence, small system sizes. Specifically, we are able to use the DMRG algorithm to access sizes up to $L=256$ with great accuracy. For the following, it is convenient to introduce the normalized energy density $\epsilon=(E - E_\mathrm{min})/(E_\mathrm{max} - E_\mathrm{min})$ with $E_\mathrm{min}$ and $E_\mathrm{max}$ the extremal eigenenergies.

\begin{figure}[!t]
    \includegraphics[width=1\columnwidth,clip]{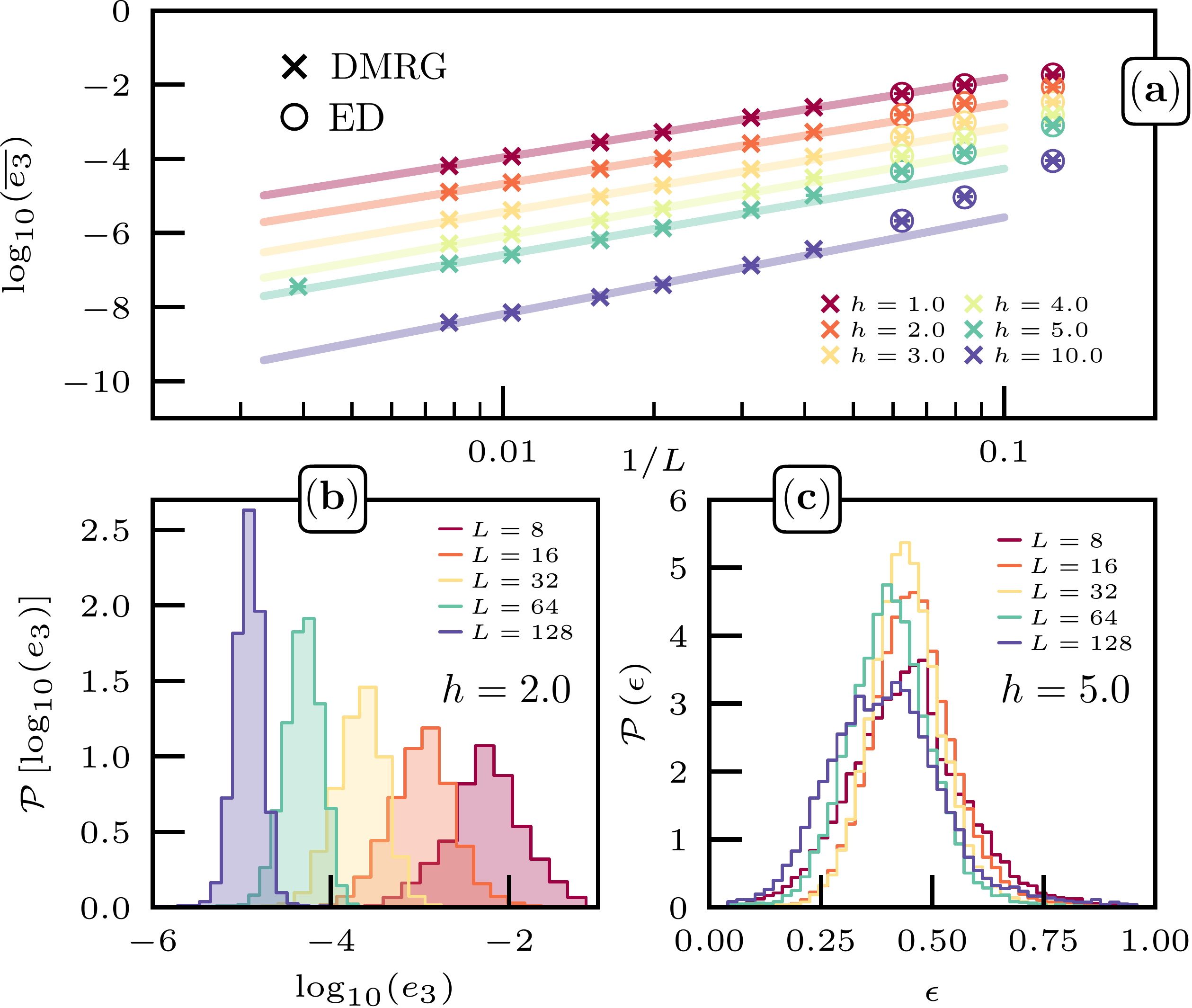}
    \caption{(Color online) (a) Scaling with the inverse system size $1/L$ of the third smallest eigenvalue $e_3$ of the covariance matrix,  corresponding to the energy variance of the input state $|\Psi_0\rangle$ with respect to the new Hamiltonian $\tilde{\mathcal{H}}$. The average is performed over $10^4$ independent samples for various disorder strengths $h$. (b) Distribution of $e_3$ at fixed disorder strength $h=2$ for different system sizes $L$. (c) Distribution of the normalized energy $\epsilon$ ($0$ and $1$ correspond respectively to the ground and most excited state) at fixed disorder strength $h=5$ of the input state $|\Psi_0\rangle$ in the new Hamiltonian spectrum $\tilde{\mathcal{H}}_3$ for different system sizes $L$.}
    \label{fig:e3_scaling}
\end{figure}

For various disorder strengths and system sizes, we compute the eigenpairs of the covariance matrix. First using ED, we always find that the eigenvalues $e_1$ and $e_2$ are, up to numerical precision, exactly zero. As expected, they are trivially associated with the initial Hamiltonian $\mathcal{H}_0$ and the (conserved) total magnetization~\cite{note}.
We now turn our attention to the third smallest eigenvalue, $e_3$, which is not strictly equal to zero. However,
it is instructive to study its scaling versus the system size $L$ for different disorder strengths $h$. Especially, since $e_3$ is nothing but the energy variance of $|\Psi_0\rangle$ with respect to a new Hamiltonian $\tilde{\mathcal{H}}$ represented by $\boldsymbol{p}_3$, we ask if it can be ``sufficiently small'' such that the input GS can correctly describe one of its eigenstates. Results are displayed in Fig.~\ref{fig:e3_scaling}\,(a), where the average value of $e_3$ over thousands of disordered samples shows a power-law scaling with the system size of the form $\propto L^{-\alpha(h)}$, with $2.15\le \alpha(h)\le 2.6$ for the values of $h$ considered~\cite{SM}. Moreover, at fixed disorder strength and for increasing system sizes, the distribution of $e_3$ is self-averaging, as shown in Fig.~\ref{fig:e3_scaling}\,(b) for $h=2$. These two observations strongly suggest that in the limit of an infinitely large system, $e_3$ will eventually go to zero and that $|\Psi_0\rangle$, the localized GS of $\mathcal{H}_0$, will also be the eigenstate of another Hamiltonian spanned by $\boldsymbol{p}_3$, dubbed $\tilde{\mathcal{H}}_3$. The relatively small values of $e_3$, even for the finite system sizes numerically available, make it possible to consider the input state as a \textit{very good} approximation of an actual eigenstate of $\tilde{\mathcal{H}}_3$ to extend our study. We further note that contrary to recently proposed DMRG-like methods for excited states~\cite{lim2016,pollmann_efficient_2016,kennes_entanglement_2016,khemani2016,yu_finding_2017,devakul_obtaining_2017,wahl_efficient_2017,villalonga2018}
where the energy variance increases with the system size $L$, our method yields a power-law decaying $\sigma[{\tilde{H}},|\Psi_0\rangle]$ with $L$.

\begin{figure}[!t]
    \includegraphics[width=1\columnwidth,clip]{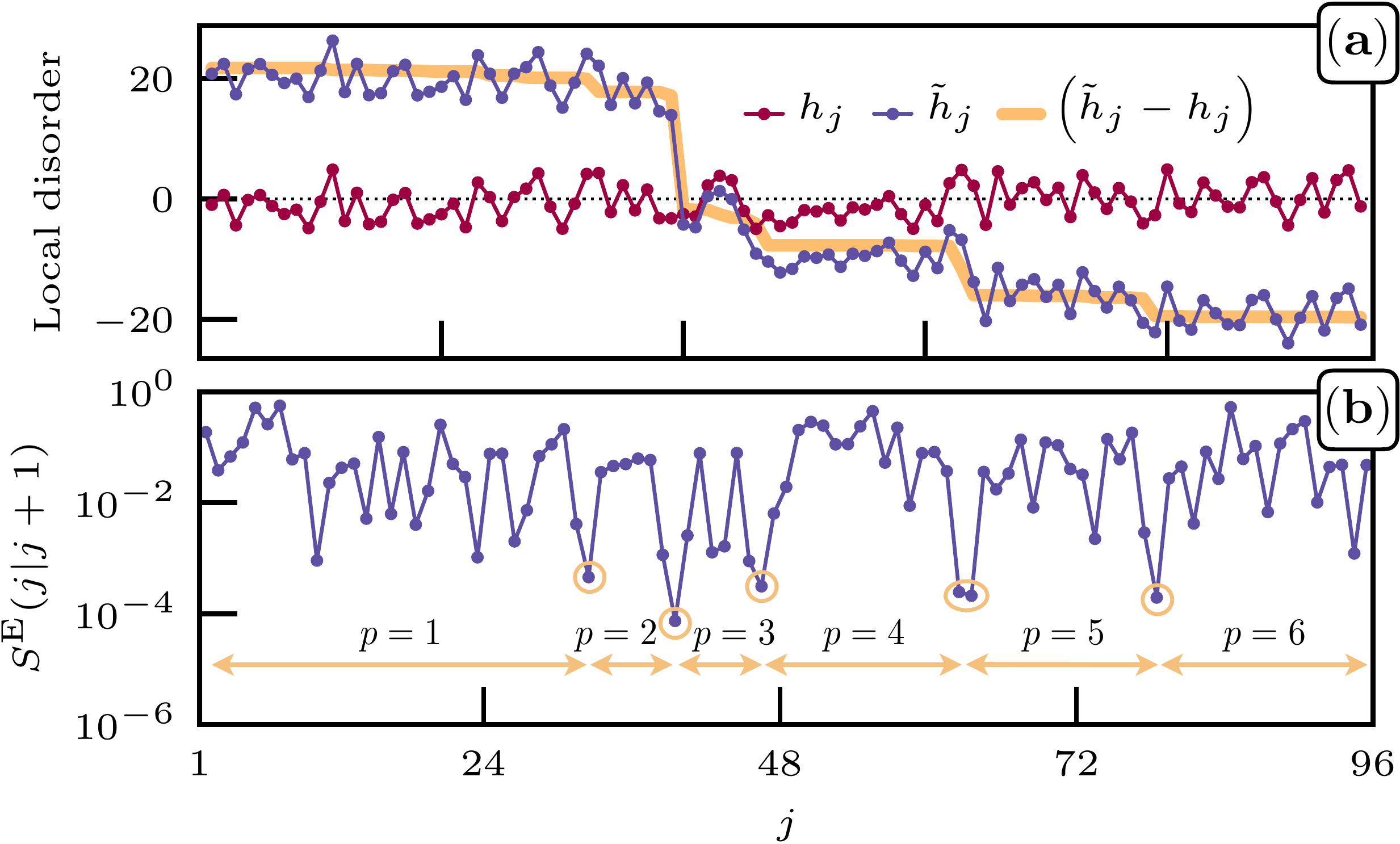}
    \caption{(Color online) Typical disordered sample of length $L=96$ with $h=5$ and an energy variance $e_3\simeq 1.348\times 10^{-7}$. (a) Local distribution in real space of the initial disorder configuration $h_j\in[-5, 5]$, the new disorder configuration $\tilde{h}_j$, and the difference between the two. (b) Bipartite von Neumann entanglement entropy $S^\mathrm{E}(j,j+1)$ between subsystems $[1,j]$ and $[j+1,L]$. The deepest minima of entanglement entropy are circled and correspond to the position of the five steps in the new disorder configuration that delimits $N_p=6$ plateaus.}
    \label{fig:hlocal_sample}
\end{figure}

The nature of $|\Psi_0\rangle$ is given by its position in the spectrum of $\tilde{\mathcal{H}}_3$, held in the normalized energy $\epsilon$. Its distribution is plotted in Fig.~\ref{fig:e3_scaling}\,(c) for $h=5$ and various system sizes, with a maximum density at high energy, $\epsilon\approx 0.5$.
Essentially, this tells us that the input GS $|\Psi_0\rangle$ is also an excited eigenstate of some other Hamiltonian, $\tilde{\mathcal{H}}_3$ and more generally that a localized GS is similar to an excited state.
One might also wonder what happens regarding the other eigenvalues of the covariance matrix, $e_j$ with $j>3$. In other words, are there more Hamiltonians, besides $\pm\mathcal{H}_0$ and now $\tilde{\mathcal{H}}_3$, for which the MBL state $|\Psi_0\rangle$ would also be an eigenstate\,? The same analysis has been performed for the other eigenpairs of the covariance matrix, with similar conclusions~\cite{SM}. Precisely we find that there actually exists a whole set of Hamiltonians $\{\tilde{\mathcal{H}}_j\}$ in the thermodynamic limit for which the input localized GS $|\Psi_0\rangle$ is an MBL excited eigenstate, classifying the MBL phenomena as \textit{a large family of ground states}.\\
\\
\textit{Inspection of the new disordered Hamiltonian.}
Focusing on the parent Hamiltonian labeled $\tilde{\mathcal{H}}_3$, it is instructive to look at its disorder configuration compared to the initial one from which $|\Psi_0\rangle$ has been computed, as shown for a typical disordered sample in Fig~\ref{fig:hlocal_sample}\,(a). In particular, computing their difference brings out the strong correlation that exists between the two. The new disorder configuration displays sharp step-like features where locally on each plateau $p$, the disorder has the same form as the original one. From $\mathcal{H}_0$ to $\tilde{\mathcal{H}}_3$, the local random fields undergo a transformation of the form $\sum_{j=1}^L h_j\longrightarrow \sum_p^{N_p}\sum_{j\in p} (h_p+h_j)$, where $h_p$ is roughly constant for a plateau $p$ of length $\ell_p$ (there are $N_p$ of them). The average number of such plateaus scales as $\overline{N_p}\propto L^\omega$ with $\omega\approx 0.7$~\cite{SM}, and the average length $\overline{\ell_p}\propto L^{1-\omega}$.

All along a given plateau $p$, the disorder configuration $\{{\tilde{h}}_j\}$ is similar to the original one $\{{{h}}_j\}$ except from a global constant shift $h_p$. Because the magnetization $S^z$ is only globally conserved and can fluctuate among different plateaus, such random shifts $h_p$ allow the state $|\Psi_0\rangle$ to have a much  higher energy. But how can $|\Psi_0\rangle$  can still be an eigenstate of $\tilde{\mathcal{H}}_3$\,? To answer this, it is crucial to observe the entanglement profile along the chain, as shown in Fig.~\ref{fig:hlocal_sample}\,(b) for the same sample as Fig.~\ref{fig:hlocal_sample}\,(a). Indeed, one can make a direct correspondence between the positions of the steps and the minima of the bipartite von Neumann entanglement entropy, defined as $S^\mathrm{E}(j|j+1)=-\sum_i\lambda_i\ln\lambda_i$, where $\lambda_i$ are the eigenvalues of the reduced density matrix of the subsystem comprised in $[1,j]$ with respect to the other part. Note that such an entanglement minima feature has also been observed  in the case of excited states~\cite{luitz2016}.

With a very small bipartite entanglement entropy between two parts of the system corresponding to almost disconnected subsystems, it is natural that the deep potential barriers of $\{{\tilde{h}}_j\}$ will occur precisely at such minima. Within each plateau, the new random fields display strong fingerprints of the initial ones since the physical properties of $|\Psi_0\rangle$ are indubitably dependent of the underlying parent Hamiltonian(s). Nevertheless, the modulation of the fields by a piecewise additive constant is what brings $|\Psi_0\rangle$ from a GS to a highly excited state.

\begin{figure}[!t]
    \includegraphics[width=1\columnwidth,angle=0,clip]{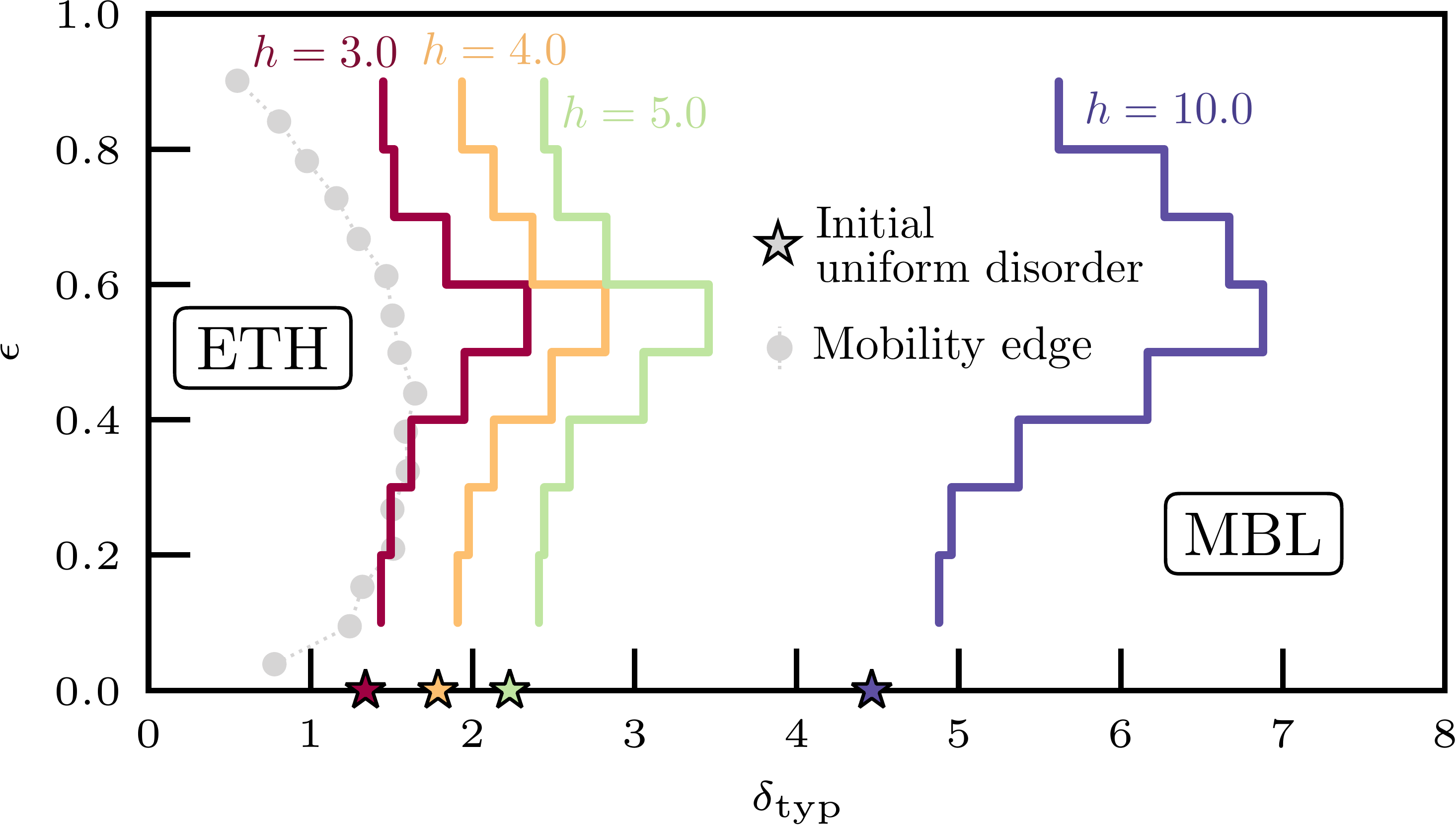}
    \caption{(Color online) Diagram of ``normalized energy density $\epsilon$ --- effective disorder strength $\delta_\mathrm{typ}$'' (see text) for the input state $|\Psi_0\rangle$ with respect to $\tilde{\mathcal{H}}_3$ for $L=128$. The local differences $\delta_i$ are considered independently over $10^4$ disordered samples to compute $\delta_\mathrm{typ}(\epsilon)$. The typical value of the initial uniform distribution $\{h_i\}$ is displayed by a star symbol on the $\epsilon=0$ line and is always smaller than for the new disorder $\{\tilde{h}_i\}$ configuration. The mobility edge in gray between the MBL and ETH regimes is taken from Ref.~\onlinecite{luitz2015} by translating the original disorder strength definition of the $x$-axis to our own, i.e., $\delta_\mathrm{typ}$.}
    \label{fig:typ_diff}
\end{figure}

In order to quantify the strength of this new disorder, we introduce the local differences $\delta_i=|{\tilde{h}}_i-{\tilde{h}}_{i+1}|$ which capture both the original randomness $h_i$ {\it and} the size of the successive jumps between plateaus. The typical value $\delta_{\rm typ}=\exp({\overline{\ln\delta_i}})$ is shown in Fig.~\ref{fig:typ_diff} for different values of the original disorder $h=3,\,4,\,5$, and $10$ in an energy-resolved diagram. There, we clearly see that the new disorder is stronger than the initial one, and we also observe an interesting dependence on $\epsilon$ which qualitatively follows the mobility edge of the original model~\cite{luitz2015}.\\
\\

\begin{figure*}[!ht]
    \includegraphics[width=2\columnwidth,clip]{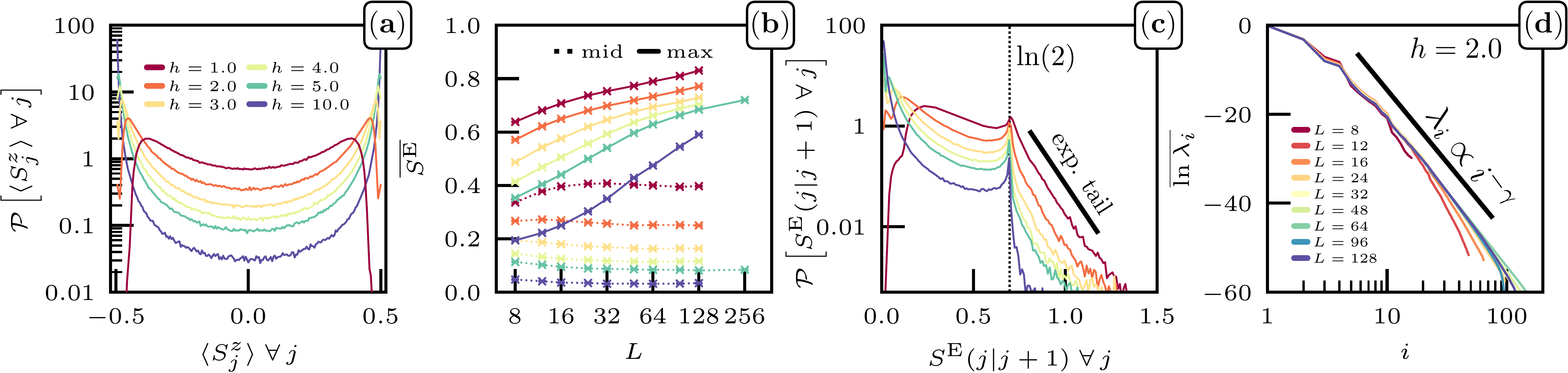}
    \caption{(Color online) Distributions and averages in this figure are computed over GS of Eq.~\eqref{eq:ham} for $10^4$ disordered samples. (a) Histogram of the local magnetizations $\langle S^z_j\rangle$ for different disorder strengths $h$ and $L=128$. (b) Average bipartite von Neumann entanglement entropy $\overline{S^\mathrm{E}}$ as a function of system size $L$ for different disorder strengths $h$. The entanglement entropy for a cut in the middle of the system is shown (dotted line) as well as the maximal entanglement entropy in the system with regard to all possible bipartitions (plain line). Note the semi-log scale. The legend is displayed in the first panel. (c) Histogram of bipartite von Neumann entanglement entropy $S^\mathrm{E}(j|j+1)$ for different disorder strengths $h$ and $L=128$. The distribution contains independently all the possible bipartitions $(j|j+1)$ of a sample with $j\in[1,L-1]$. The legend is displayed in the first panel. (d) Average of the logarithm of the entanglement spectrum values $\{\lambda_i\}$ (sorted in descending order) for a bipartition in the middle of the system as a function of their indices $i$. A disorder strength $h=2$ is considered for various system sizes.}
    \label{fig:dist_bg_mbl}
\end{figure*}

\noindent\textit{Similarity between localized GS and MBL.} To complete our study, we now argue that given any set of physical measurements done on a localized eigenstate, ground and excited states appear to be barely indistinguishable. In particular, we show that local magnetization, bipartite von Neumann entanglement entropy and the entanglement spectrum properties of MBL eigenstates are similar for ground and excited states.

In the GS $|\Psi_0\rangle$ of Eq.~\eqref{eq:ham}, while quantum fluctuations prevent an exact alignment of the magnetic moments with the random field, the spins will nevertheless locally follow the field pattern in order to minimize the energy of the system.
This results in strongly polarized spins, with typically $|\langle S^z_j\rangle|-0.5\ll 1$, as visible in the histogram of local magnetizations in Fig.~\ref{fig:dist_bg_mbl}\,(a). This distribution is similar in many ways to MBL excited states with a double peak structure~\cite{khemani2016,lim2016}, and a density $\mathcal{P}\left[\langle S^z\rangle=0\right]$ decreasing with $h$. At high energy, it is a fingerprint of ergodicity breaking, where the single-site distribution is totally different from a thermal distribution, unlike the ETH phase at a smaller disorder strength.

Another characteristic property of MBL excited states is their area law for the entanglement entropy~\cite{bauer2013,kjall2014,luitz2015,lim2016,khemani2017},
best known to be the hallmark of the GS of \textit{any} generic short-range Hamiltonian~\cite{hastings2007,eisert2010,laflorencie2016}. In Fig.~\ref{fig:dist_bg_mbl}\,(b), we show as dotted lines the average value of the bipartite von Neumann entropy $\overline{S^\mathrm{E}}$ (with a cut in the middle of the system) which clearly saturates to an area law. Perhaps more interestingly, one can also study the ``optimal cut'' entanglement entropy targeting the maximal entropy over all possible bipartitions in a given sample. Its mean value is plotted with plain lines in Fig.~\ref{fig:dist_bg_mbl}\,(b) where a logarithmic growth $\propto \ln L$ is observed, in agreement with Ref.~\onlinecite{bauer2013} for the MBL regime, contrasting with the strict area law obtained for the middle chain cut. Such a peculiar logarithmic violation of a strict area law can be understood from the histogram plotted in Fig.~\ref{fig:dist_bg_mbl}\,(c) where three main regions are visible, again in quantitative agreement with MBL~\cite{bauer2013,luitz2016}: (i) a maximum at very small values signaling that most of the cuts display tiny entanglement~; (ii) a secondary maximum at $S^\mathrm{E}=\ln 2$ which comes from a local singlet formation where the random fields are locally small, such a peak being slowly suppressed when $h$ increases ; and (iii) an exponential tail at a larger value of $S^\mathrm{E}$ which traces back (exponentially) rare events where disorder is locally weaker over a finite length, yielding an entanglement much larger than the average. These rare regions, whose density $\propto\exp(-aS^\mathrm{E})$, leads to an optimal cut entanglement $\propto\ln L$, as already understood for the MBL regime in Ref.~\onlinecite{bauer2013}.

One sees that the entanglement properties of a localized GS $|\Psi_0\rangle$ are quantitatively very comparable to MBL physics. Furthermore, one can also study the entanglement spectrum, corresponding to the eigenvalues $\{\lambda_i\}$ of the reduced density matrix. Already studied in the context of MBL~\cite{yang_two-component_2015,geraedts_many-body_2016,serbyn_power-law_2016,pietracaprina_entanglement_2017,geraedts_characterizing_2017,gray_many-body_2018}, a power-law distribution of the form ${\overline{\ln \lambda_i}}\propto i^{-\gamma(h)}$ was found~\cite{serbyn_power-law_2016}, contrasting with flatness in the ETH case~\cite{yang_two-component_2015} and exponential decay for gapped GS~\cite{chung_density-matrix_2001}. Here, we strikingly observe a power-law behavior~\cite{SM} for the entanglement levels, as plotted in Fig.~\ref{fig:dist_bg_mbl}\,(d), showing again similar behavior between localized GS and MBL physics.\\
\\
\noindent\textit{Discussions and conclusions.} Using large-scale numerical simulations, we have found that in the presence of disorder a single eigenstate does not uniquely encodes its underlying Hamiltonian, since a localized many-body GS is a very good approximation of an eigenstate of another Hamiltonian that only differs by its local disorder configuration from the original one. Precisely, with respect to the new Hamiltonian, it corresponds to a highly excited state, even though all its properties are by definition those of a GS. This connects localized GS to MBL physics of highly excited states. In this sense, we have complemented our study showing that given any set of physical measurements performed on a localized eigenstate, ground and excited states appear scarcely indiscernible.
We believe that this ``eigenstate-to-Hamiltonian construction'' method provides an interesting alternative to other variational approaches based on building matrix-product states for excited states.

An interesting continuation of this work would be to extend it to higher dimensions, although more numerically challenging. In particular, we believe that it would allow one to tackle the MBL phenomena in two dimensions where only a few theoretical studies are available~\cite{chandran_many-body_2016,lev_slow_2016,wahl_signatures_2017,thomson_time_2018,bertoli_finite_2017}, despite a recent experimental observation~\cite{choi_exploring_2016}.\\

\begin{acknowledgments}
    \noindent\textit{Acknowledgments.} We are grateful to F. Alet, D. J. Luitz, and N. Mac\'e for interesting comments. We acknowledge support of the French ANR programs BOLODISS (Grant No. ANR-14-CE32-0018) and THERMOLOC (Grant No. ANR-16-CE30-0023-02). This work was also funded by R\'egion Midi-Pyr\'en\'ees and by the U.S. Department of Energy, Office of Science, Office of Basic Energy Sciences, Materials Sciences and Engineering Division under Contract No. DE-AC02-05-CH11231 through the Scientific Discovery through Advanced Computing (SciDAC) program (KC23DAC Topological and Correlated Matter via Tensor Networks and Quantum Monte Carlo). The numerical simulations were performed using HPC resources from GENCI (Grants No. x2017050225, No. A0010500225, and No. A0030500225) and CALMIP. The calculations involving the DMRG algorithm were done using the ITensor library~\footnote{ITensor library, \href{http://itensor.org}{http://itensor.org}.}.
\end{acknowledgments}

\clearpage

\appendix
\section{Supplemental material to ``\textit{Many-body localization as a large family of localized ground states}''}

\begin{figure}[!h]
    \includegraphics[width=1\columnwidth,clip]{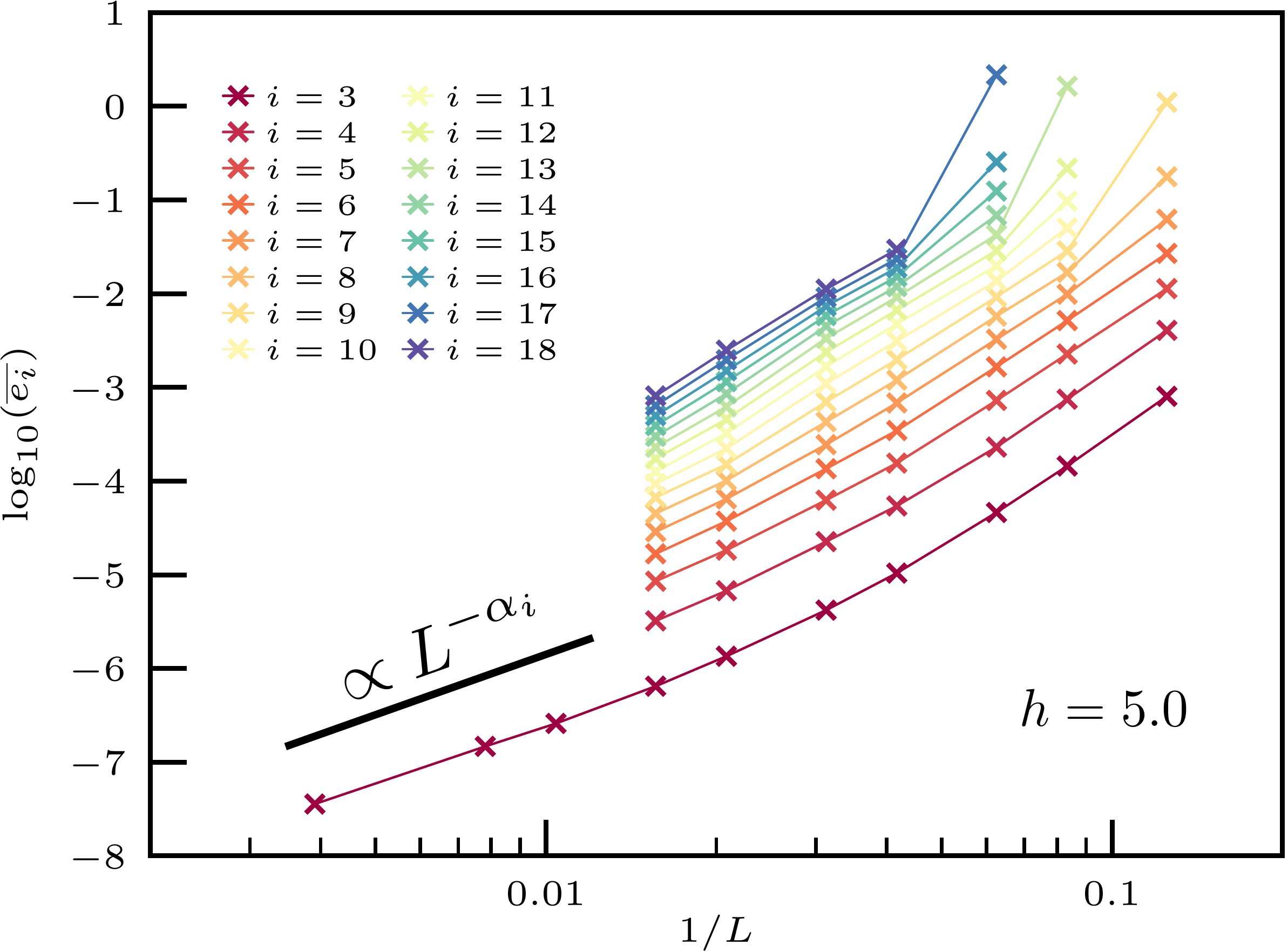}
    \caption{(Color online) Scaling with the inverse system size $1/L$ of the smallest eigenvalues $e_j$ of the covariance matrix. They are sorted in ascending order and correspond to the energy variance of the input state $|\Psi_0\rangle$ with respect to a new Hamiltonian $\tilde{\mathcal{H}}_j$. The eigenvalues corresponding to $j=1$, $2$ have been omitted since they are associated to the initial Hamiltonian $\pm\mathcal{H}_0$ from which $|\Psi_0\rangle$ is by definition an eigenstate. The disorder strength of the initial Hamiltonian is $h=5.0$. The average is performed over $10^4$ independent samples for $j=3$ and $10^3$ independent samples for $j\geq 4$.}
    \label{fig:other_e}
\end{figure}

\noindent{\bf (i) Smallest eigenvalues of the covariance matrix}\\
The third eigenvalue decays rapidly with system size $e_3\propto L^{-\alpha_3(h)}$. In Table~\ref{tab:1} below we give the estimates for $\alpha_3(h)$. Interestingly, not only the third eigenvalue $e_3$ decays, but  most of them will eventually vanish at the thermodynamic limit. This is illustrated in Fig.~\ref{fig:other_e}, and in Table~\ref{tab:2}

\begin{table}[!h]
\begin{center}
\begin{tabular}{c||c|c|c|c|c|c}
$h$&1&2&3&4&5&10\\
\hline
$\alpha_3$&2.15(3) & 2.16(4) &2.28(7) & 2.36(9) & 2.3(1) & 2.6(1)
\end{tabular}
\end{center}
\label{tab:1}
\caption{Decay exponent $\alpha_3$ of the third eigenvalue $e_3\propto L^{-\alpha_3(h)}$ of the covariance matrix as a function of disorder  $h$.}
\end{table}
\begin{table}[!h]
\begin{center}
\begin{tabular}{c||c|c|c|c|c|c}
$i$&3&4&5&6&7&8\\
\hline
$\alpha_i$&2.3(1) &2.9(1) &2.97(8)& 3.12(9)&3.2(1)&3.4(2)
\end{tabular}
\end{center}
\label{tab:2}
\caption{$\alpha_i$ of the i$^{\rm th}$ eigenvalue $e_i\propto L^{-\alpha_i(h)}$ for   $h=5$.}
\end{table}

\begin{figure}[!h]
    \includegraphics[width=1\columnwidth,clip]{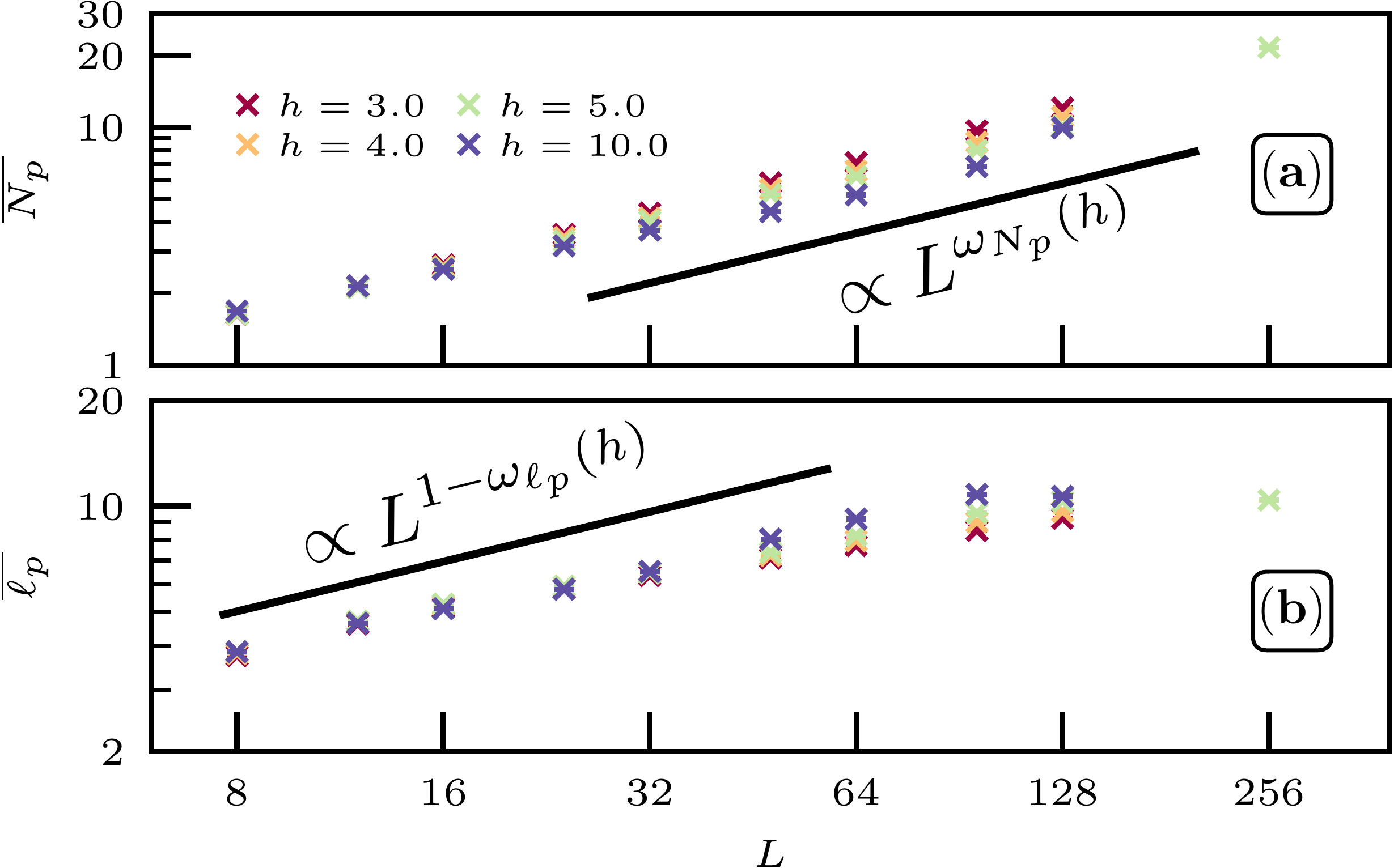}
    \caption{(Color online) Average (a) plateaus number $\overline{N_p}$ and (b) length $\overline{\ell_p}$ in the new disorder configuration $\{\tilde{h}_j\}$ as function of the system size $L$ for various disorder strengths $h$. The average is performed over $10^4$ independent samples. We define two different plateaus by a step of height $\geq 1$ in between.}
    \label{fig:plateaus}
\end{figure}

\noindent {\bf{(ii) Properties of the new disorder configuration}}\\
The new disorder configuration $\{{\tilde{h}}_j\}$ of the parent Hamiltonian ${\tilde{\cal{H}}}_3$ resembles the original one $\{h_j\}$ plus an additional stair-like structure made of plateaus. In Fig.~\ref{fig:plateaus} we show (a) how the average number of plateaus ${\overline{N_p}}$ scales with the length $L$, and (b) the average plateau length ${\overline{\ell_p}}$. They both scale with an exponent $\omega<1$ such that ${\overline{N_p}}\propto L^{\omega_{N_p}}$ and ${\overline{\ell_p}}\propto L^{1-\omega_{\ell_p}}$. Results are shown in Fig.~\ref{fig:plateaus} and the estimates for $\omega$ are displayed in Table~\ref{tab:3} where one sees that they agree within error bars to the value $\omega\approx 0.7$, with no clear $h$ dependence.\\
\begin{table}[!h]
\begin{center}
\begin{tabular}{c||c|c|c|c}
$h$&3&4&5&10\\
\hline
$\omega_{N_p}$&0.732(5) &0.700(8)&0.86(5)&0.67(5)\\
$\omega_{\ell_p}$&0.70(2)&0.69(1)&0.73(3)&0.62(3)
\end{tabular}
\end{center}
\label{tab:3}
\caption{Exponents $\omega$ governing the plateaus structure of the new disorder configuration $\{{\tilde{h}}_j\}$.}
\end{table}

\newpage

\begin{figure}[!h]
    \includegraphics[width=1\columnwidth,clip]{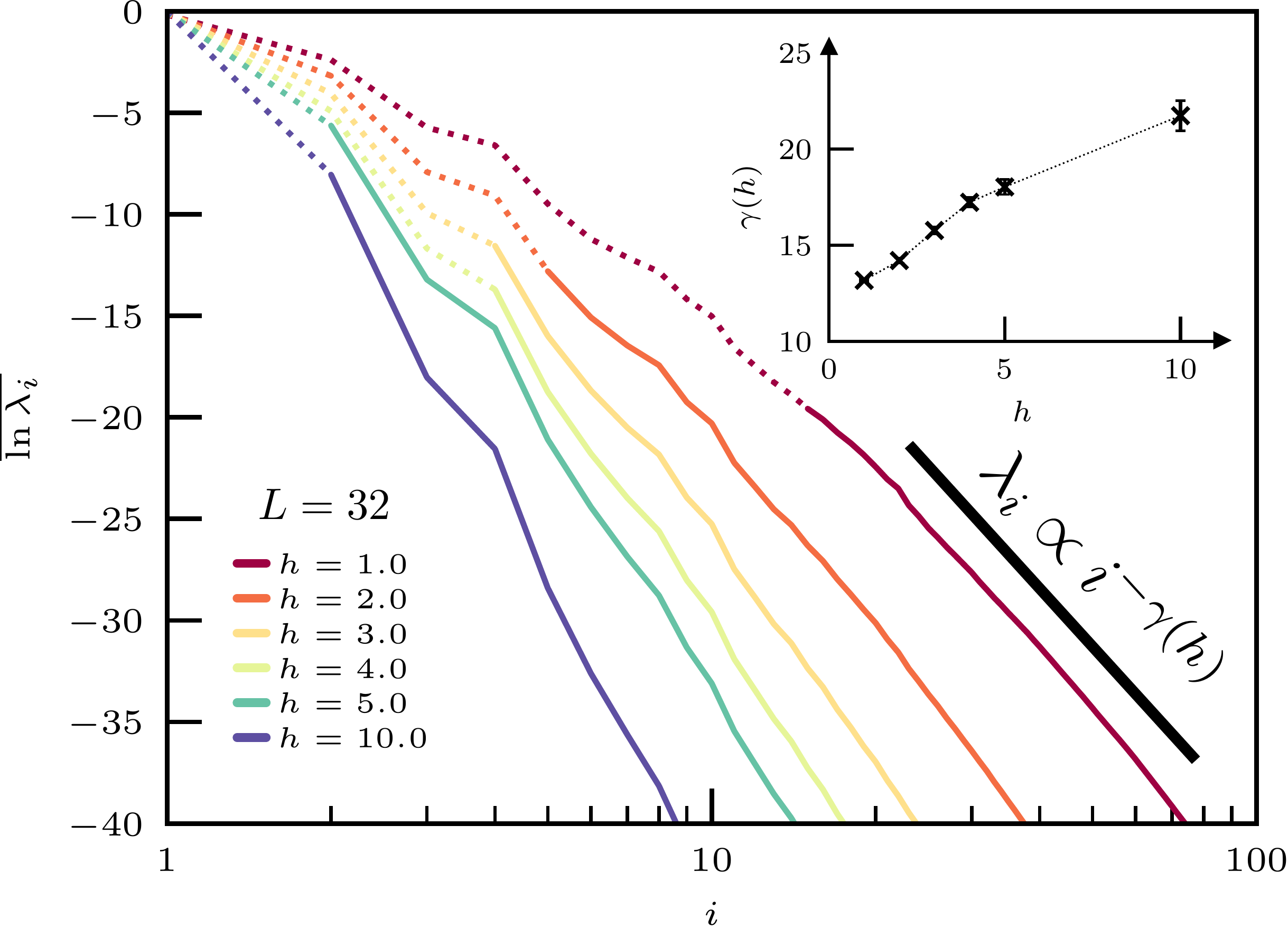}
    \caption{(Color online) Average of the logarithm of the entanglement spectrum values $\{\lambda_i\}$ (sorted in descending order) for a bipartition in the middle of the system as a function of their indices $i$. A system size $L=32$ is considered for various disorder strengths $h$. A power-law dependence of the form $\overline{\mathrm{ln}\lambda_i}\propto i^{-\gamma(h)}$ is observed and a fit is performed over the plain line region. The exponent $\gamma(h)$ versus the disorder strength is reported in the inset.}
    \label{fig:es_scaling}
\end{figure}

\noindent{\bf (iii) Power-law scaling of the entanglement spectrum}\\
The entanglement spectrum is shown in Fig.~\ref{fig:es_scaling} for $L=32$ and various values of disorder strengths. The power-law decay $\lambda_i\propto 1/i^{\gamma(h)}$ is clearly visible, with an exponent $\gamma$ which varies with $h$. The behavior $\gamma(h)$ is shown in the inset of Fig.~\ref{fig:es_scaling} where one sees that it can take quite large values deep in the localized regime.

\begin{thebibliography}{73}
\bibitem [{not()}]{note}%
  \BibitemOpen
  \href@noop {} {}\bibinfo {note} {The eigenvectors of the covariance matrix
  form an orthonormal basis with
  $\boldsymbol{p}_i\cdot\boldsymbol{p}_j=\delta_{ij}$. The normalization means
  that each set of parameters contained in $\boldsymbol{p}_i$ fulfills
  $\tilde{J}^2+\sum_j \tilde{h}_j^2=1$ whereas $J=1$ is considered in the
  initial Hamiltonian. We therefore enforce $\tilde{J}=J=1$ and rescale the
  fields $\tilde{h}_j$ accordingly to have comparable energy scales between the
  initial and the parent Hamiltonians. Concerning the double degeneracy of the
  zero eigenvalues of the covariance matrix, $e_1=e_2=0$, it comes from the
  total magnetization $S^z_\mathrm{tot}$ conservation of the initial
  Hamiltonian $\mathcal{H}_0$: Any linear combination of these two operators is
  a conserved quantity and the two-dimensional kernel of the covariance matrix
  $\mathsf{C}$ is spanned by these two different operators. The position
  $\epsilon$ of the input state for the parent Hamiltonians defined by
  $\boldsymbol{p}_1$ and $\boldsymbol{p}_2$ is equal to $0$ and
  $1$.}\BibitemShut {Stop}%
\end{thebibliography}
\end{document}